\documentstyle[11pt,epsf]{article}
 1
 1
 1

\renewcommand{\thefootnote}{\fnsymbol{footnote}}
\def\lsim{\mathrel{\raise.3ex\hbox{$<$\kern-.75em\lower1ex\hbox{$\sim$}}}}
\def\gsim{\mathrel{\raise.3ex\hbox{$>$\kern-.75em\lower1ex\hbox{$\sim$}}}}
\begin{document}
\noindent
\thispagestyle{empty}
\renewcommand{\thefootnote}{\fnsymbol{footnote}}
\begin{flushright}
{\bf TTP96-04}\footnote[1]{The complete postscript file of this
preprint, including figures, is available via anonymous ftp at
www-ttp.physik.uni-karlsruhe.de (129.13.102.139) as /ttp96-04/ttp96-04.ps 
or via www at http://www-ttp.physik.uni-karlsruhe.de/cgi-bin/preprints/.}\\
{\bf MPI/PhT/96-10}\\
{\bf DPT/96/12}\\
{\bf hep-ph/9603313}\\
{\bf March 1996}\\
\end{flushright}
\begin{center}
  \begin{large}
DOUBLE BUBBLE CORRECTIONS TO HEAVY QUARK PRODUCTION\footnote[3]{
     Work supported by BMFT under Contract 056KA93P6, 
     DFG under Contract Ku502/6-1 and INTAS under Contract INTAS-93-0744.}\\
  \end{large}
  \vspace{0.5cm}

\begin{large}
 K.G.~Chetyrkin$^{a}$,
 A.H. Hoang$^{b}$\footnote[2]{Supported by 
              Graduiertenkolleg Elementarteilchenphysik, Karlsruhe.},
 J.H.~K\"uhn$^{b}$, 
 M.~Steinhauser$^{b}$\footnotemark[2],
 T.~Teubner$^{c}$
\end{large}
\begin{itemize}\centering
\item[$^a$]
   Max-Planck-Institut f\"ur Physik, Werner-Heisenberg-Institut,\\
   F\"ohringer Ring 6, 80805 Munich, Germany.
\item[$^b$]
   Institut f\"ur Theoretische Teilchenphysik\\ 
   Universit\"at Karlsruhe, Kaiserstr. 12,    Postfach 6980,
   D-76128 Karlsruhe, Germany\\  
\item[$^c$]
   Department of Physics\\
   University of Durham,   Durham, DH1 3LE, UK
\end{itemize}

  \vspace{0.7cm}
  {\bf Abstract}\\
\vspace{0.3cm}
\renewcommand{\thefootnote}{\arabic{footnote}}
\addtocounter{footnote}{-2}
\noindent
\begin{minipage}{15.0cm}
\begin{small}
Second order ${\cal O}(\alpha_s^2)$ corrections to the heavy quark 
production cross-section due to massless quarks and coloured scalars  
are calculated for all energies above threshold. Based on the method
introduced in this letter also the gauge non-invariant second order
corrections due to the pure gluonic selfenergy insertion and a certain
class of ${\cal{O}}(\alpha_s^3)$ and
${\cal{O}}(\alpha_s^4)$ corrections are
determined. For the special choice of the gauge parameter, $\xi=4$, the
leading threshold and high energy behaviour of the pure second order
gluonic corrections to the cross-section are governed by the
gluonic self energy insertion. 
\end{small}
\end{minipage}
\end{center}
\vspace{1.2cm}
%
%
%
\noindent
\section{Introduction}
\label{sectionintroduction}
As a consequence of the high precision measurements at LEP
the total cross-section for quark anti-quark production in
$e^+e^-$-annihilation has been subject to extensive studies
during the past years. Whereas the ${\cal{O}}(\alpha_s)$ corrections are
known for all energy and mass values~\cite{KaeSab55}, the
complete ${\cal{O}}(\alpha_s^2)$ corrections are only known in the high
energy expansion including terms up to ${\cal{O}}(M^4/s^2)$, where $s$
denotes the c.m. energy and $M$ the mass of the produced quarks,
see~\cite{CheKueKwiRep}.
However, in view of future experiments ($\tau$-charm-, B-factory, NLC)
where quark anti-quark pairs will be produced near their production
threshold, the knowledge of the complete ${\cal{O}}(\alpha_s^2)$
corrections for all mass and energy assignments is desirable.
Analytical formulae are of particular importance because they provide
important cross-checks for approximation methods which can be applied
even where an analytical evaluation seems to be impossible.
A closed analytical expression  for the vector-current induced
${\cal{O}}(\alpha_s^2)$ corrections due to a massless fermion pair has
been published in~\cite{HoaKueTeu95}.
The complete vector-current induced 
gluonic contribution were obtained recently
in~\cite{CheKueSte95} using Pad\'e approximation methods.
\par
In this letter a refinement and an extension of the results obtained
in \cite{HoaKueTeu95} are presented.
In section~2 the
second order corrections to the total 
cross-section of massive fermion pair production due to 
radiation of a light secondary
fermion pair from the final state will be presented. The 
framework of on-shell renormalized QED will be employed
with the fine structure constant 
as the expansion parameter of the perturbation series. 
The result will be
parametrized in terms of moments in which all information on the vacuum
polarization due to the light fermion pair is encoded.
As a consequence the corresponding corrections due to a pair of light
scalar particles can be easily determined.
In the framework of a supersymmetric model this result
can be applied to light sfermion radiation in $t\bar t$ production. 
The concept of the moments even allows for a determination of those
fermionic ${\cal{O}}(\alpha^3)$ corrections to the total inclusive massive
fermion pair production cross-section which 
originate from the insertion of the
one-loop corrected vacuum polarization into the photon line.
The remaining logarithm of the square of the small mass
devided by the c.m. energy can be absorbed by employing the running
coupling. 
In section~3 the transition to QCD will be performed. Based on
the results of section~2 the gauge non-invariant
${\cal{O}}(\alpha_s^2)$ contributions due to the gluonic self-energy
will be presented. The result will be examined for the special gauge
$\xi=4$, where the leading contributions in the threshold region as well
as for high energies are determined by the self-energy insertion.
In section~4, finally, we apply the concept of moments to determine
a certain class of fermionic ${\cal{O}}(\alpha_s^4)$ corrections without 
referring back to the corresponding QED result. 
This calculation constitutes an
example how the method of moments can be used to fix some even
higher order QCD corrections.
\section{Light Fermionic and Scalar Second Order Corrections}
\label{sectionlightfermionic}
\begin{figure}[ht]
 \begin{center}
 \begin{tabular}{ccc}
   \epsfxsize=3.5cm
   \leavevmode
   \epsffile[170 270 420 520]{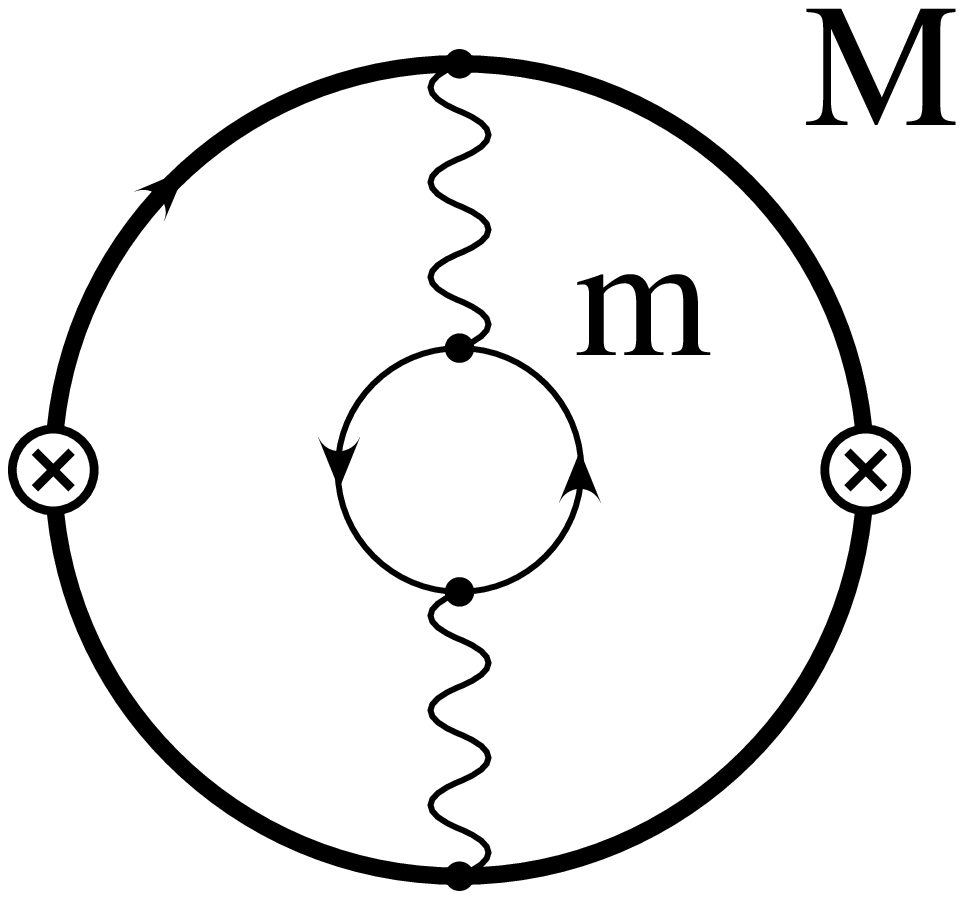}
   & \hspace{10ex} &
   \epsfxsize=3.5cm
   \leavevmode
   \epsffile[170 270 420 520]{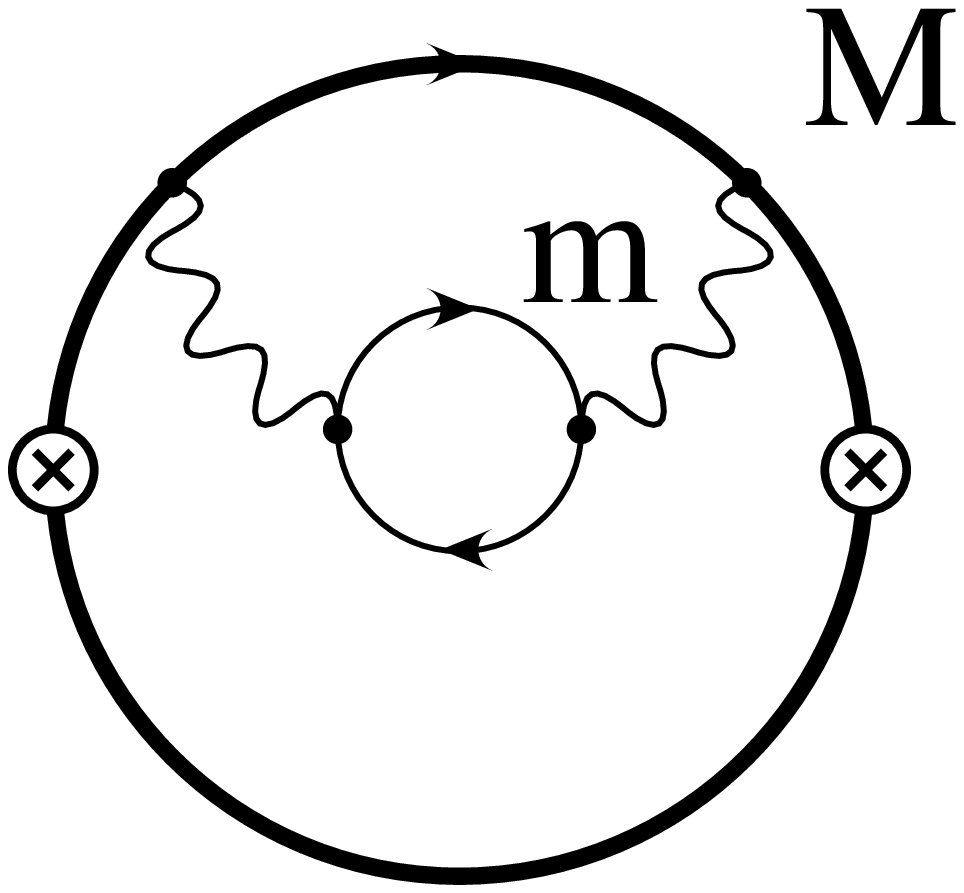}
 \end{tabular}
 \caption{\label{figferm} Fermionic double bubble diagrams.}
 \end{center}
\end{figure}

We consider the vector-current induced inclusive 
cross-section for the production of a fermion anti-fermion pair, 
$F\bar F$ (with mass $M$), normalized to the point cross-section,
\begin{equation}
R_{F\bar F}\, = \, \frac{\sigma(e^+e^-\to\gamma^*\to F\bar F\ldots)}
                      {\sigma_{pt}}\,,
\qquad\qquad
  \sigma_{pt}\, = \, \frac{4\,\pi\,\alpha^2}{3\,s}
\end{equation}
for arbitrary c.m. energy, $\sqrt{s}\ge 2\,M$. 
We only discuss final state corrections and restrict 
ourselves in this section to on-shell
renormalized QED.
The perturbative series of $R_{F\bar F}$ reads
\begin{equation}
R_{F\bar F}\, = \, r^{(0)} + 
               \,\Big(\frac{\alpha}{\pi}\Big)\,r^{(1)} +
               \,\Big(\frac{\alpha}{\pi}\Big)^2\,r^{(2)} +\, ...\,,
\label{rexpanded}
\end{equation}
where $\alpha=1/137$ is the fine structure constant defined in the
Thomson limit. The Born and first order~\cite{KaeSab55} contributions 
are well known:
\begin{eqnarray}
r^{(0)} & = & \frac{\beta}{2}\,(3-\beta^2)\,,\nonumber\\[2mm]
r^{(1)} & = &
\frac{\left( 3 - {\beta^2} \right) \,\left( 1 + {\beta^2} \right) }{2
    }\,\bigg[\, 2\,\mbox{Li}_2(p) + \mbox{Li}_2({p^2}) + 
     \ln p\,\Big( 2\,\ln(1 - p) + \ln(1 + p) \Big) 
      \,\bigg] \,\nonumber\,\\ 
 & & \mbox{} - 
  \beta\,( 3 - {\beta^2} ) \,
   \Big( 2\,\ln(1 - p) + \ln(1 + p) \Big)  - 
  \frac{\left( 1 - \beta \right) \,
     \left( 33 - 39\,\beta - 17\,{\beta^2} + 7\,{\beta^3} \right) }{16}\,
   \ln p\,\nonumber\,\\ 
 & & \mbox{} + 
  \frac{3\,\beta\,\left( 5 - 3\,{\beta^2} \right) }{8}
\,,
\end{eqnarray}
where
\[  p \, = \, \frac{1-\beta}{1+\beta}\,,\qquad\,
    \beta \, = \, \sqrt{1-4\,x}\,,\qquad\,
    x \, = \, \frac{M^2}{s}\,.  \]
For the convenience of the reader and for later reference we also present
the corresponding expansions near threshold, $\beta\equiv\sqrt{1-4\,x}\to
0$, and at high energies, $x\to 0$.
\begin{eqnarray}
r^{(0)} & \stackrel{\beta\to 0}{\longrightarrow} & \frac{3}{2}\,\beta 
 \,+\,{\cal{O}}(\beta^3)
\,,\nonumber \\[2mm]
r^{(1)} & \stackrel{\beta\to 0}{\longrightarrow} & 
 \frac{9}{2}\,\zeta(2) - 6\,\beta + 3\,\zeta(2)\,{\beta^2}
 \,+\,{\cal{O}}(\beta^3)
\,,\nonumber \\[2mm]
r^{(0)} & \stackrel{x\to 0}{\longrightarrow} & 
   1 - 6\,x^2 - 8\,x^3
 \,+\,{\cal{O}}(x^4)
\,,\nonumber \\[2mm]
r^{(1)} & \stackrel{x\to 0}{\longrightarrow} & 
\frac{3}{4} + 9\,x + \Big( \frac{15}{2} - 18\,\ln x \Big) \,{x^2} 
- \frac{4}{9}\,\Big( 47 + 87\,\ln x \Big) \,{x^3}
 \,+\,{\cal{O}}(x^4)
\,.
\end{eqnarray}
The contribution to $r^{(2)}$
arising from a light fermion anti-fermion pair with mass $m$ which 
corresponds to the
sum of all possible cuts of the current-current correlator diagrams
depicted in Fig.~\ref{figferm} is denoted by $r^{(2)}_f$. 
\par
The contribution where the light fermion pair is primarily produced 
is treated in~\cite{HoaJezKueTeu94} and
will not be discussed in the following.
Due to Furry's theorem the corresponding interference terms 
(singlet contributions) vanish identically,  if vector-currents are
considered. The result for $r^{(2)}_f$ can be written in the
form ($x=f$)
\begin{equation}
r^{(2)}_x \, = \,
\left\{\,
-\frac{1}{3}\Big[\,R_\infty^x\,\ln\frac{m^2}{s} - R_0^x \,\Big]\,r^{(1)}
+ R_\infty^x\,\delta^{(2)}  
  \,\right\}\,,
\label{mainformula}
\end{equation}
where the moments $R_\infty^f$ and $R_0^f$ are constants which
are uniquely determined by the vacuum polarization due to the
light fermion pair
\begin{equation}
\Pi_{light}^x(q^2) \, \stackrel{m^2\to 0}{\longrightarrow} \,
 -\frac{\alpha}{3\,\pi}\,\Big[\,
   R_\infty^x\,\ln\frac{-q^2}{4\,m^2} + R_0^x
   \,\Big]\,.
\label{vacuumpol}
\end{equation}
The combination
$\alpha^2 R_\infty^f/(3\,\pi)$ is the second order contribution to the
QED running coupling $\beta$-function, whereas $R_0^f$ is the zeroth
momentum of the vacuum polarization due to the light fermion pair, 
defined in~\cite{KniKraKueStu88}
\begin{eqnarray}
R_\infty^f & = & R_{f\bar f}(\infty) \, = \, 1,
\nonumber \\
R_0^f & = & 
      \int\limits_{4\,m^2}^{\infty}\,\frac{d s}{s}\,
         \left[R_{f\bar f}(s)- R_\infty^f \right]
\, = \, -\,\frac{5}{3}+\ln 4. 
\end{eqnarray}
$R_{f\bar f}$ is the normalized Born cross-section for pair production
of the light fermion anti-fermion pair with mass $m$.
The function $\delta^{(2)}$ has already been presented 
in~\cite{HoaKueTeu95}
in a somewhat different form
\begin{eqnarray}
\lefteqn{
\delta^{(2)} \, = \,
- \,\frac{\left( 3 - {{\beta }^2} \right) \,
       \left( 1 + {{\beta }^2} \right) }{6}\,\times\,}\nonumber\,\\ 
 & & \mbox{}\,
     \qquad\,\bigg\{\,\mbox{Li}_3(p) - 2\,\mbox{Li}_3(1 - p) - 
       3\,\mbox{Li}_3({p^2}) - 4\,\mbox{Li}_3\Big({p\over {1 + p}}\Big) - 
       5\,\mbox{Li}_3(1 - {p^2}) + 
       \frac{11}{2}\,\zeta(3)\,\nonumber\,\\ 
 & & \mbox{}\,\qquad + 
       \mbox{Li}_2(p)\,\ln\Big(\frac{4\,\left( 1 - {{\beta }^2} \right) }{
          {{\beta }^4}}\Big) + 2\,\mbox{Li}_2({p^2})\,
        \ln\Big(\frac{1 - {{\beta }^2}}{2\,{{\beta }^2}}\Big) + 
       2\,\zeta(2)\,\bigg[\, \ln(p) - 
          \ln\Big(\frac{1 - {{\beta }^2}}{4\,\beta }\Big) \,\bigg] 
      \,\nonumber\,
        \\ 
 & & \mbox{}\,\qquad - 
       \frac{1}{6}\,\ln\Big(\frac{1 + \beta }{2}\Big)\,
        \bigg[\, 36\,\ln(2)\,\ln(p) - 44\,\ln^2(p) + 
          49\,\ln(p)\,\ln\Big(\frac{1 - {{\beta }^2}}{4}\Big) + 
          \ln^2\Big(\frac{1 - {{\beta }^2}}{4}\Big) \,\bigg] \,\nonumber\,
        \\ 
 & & \mbox{}\,\qquad - 
       \frac{1}{2}\,\ln p\,\ln \beta\,
        \bigg[\, 36\,\ln(2) + 21\,\ln(p) + 16\,\ln(\beta ) - 
          22\,\ln(1 - {{\beta }^2}) \,\bigg]  \,\bigg\} \,\nonumber\,
     \\ 
 & & \mbox{}   + 
  \frac{1}{24}\,\bigg\{ \,
      ( 15 - 6\,{{\beta }^2} - {{\beta }^4} ) \,
      \Big( \mbox{Li}_2(p) + \mbox{Li}_2({p^2}) \Big)  + 
     3\,( 7 - 22\,{{\beta }^2} + 7\,{{\beta }^4} ) \,
      \mbox{Li}_2(p)\,\nonumber\,\\ 
 & & \mbox{}\,\qquad - 
     ( 1 - \beta  ) \,
      ( 51 - 45\,\beta  - 27\,{{\beta }^2} + 5\,{{\beta }^3} ) \,
      \zeta(2)\,\nonumber\,\\[2mm] 
 & & \mbox{}\,\qquad + 
     \frac{\left( 1 + \beta  \right) \,
        \left( -9 + 33\,\beta  - 9\,{{\beta }^2} - 15\,{{\beta }^3} + 
          4\,{{\beta }^4} \right) }{\beta }\,\ln^2 p\,\nonumber\,
      \\ 
 & & \mbox{}\,\qquad + 
     \bigg[ \,( 33 + 22\,{{\beta }^2} - 7\,{{\beta }^4} ) \,
         \ln 2 - 10\,( 3 - {{\beta }^2} ) \,
         ( 1 + {{\beta }^2} ) \,\ln \beta \,
   \nonumber\,    \\ 
 & & \mbox{}\,\qquad\,\qquad\,\qquad   - 
        ( 15 - 22\,{{\beta }^2} + 3\,{{\beta }^4} ) \,
         \ln\Big(\frac{1 - {{\beta }^2}}{4\,{{\beta }^2}}\Big)\,\bigg] \,
      \ln p\,\nonumber\,\\ 
 & & \mbox{}\,\qquad + 
     2\,\beta \,( 3 - {{\beta }^2} ) \,
   \ln\Big(\frac{4\,\left( 1 - {{\beta }^2} \right) }{{{\beta }^4}}\Big)\,
   \bigg[\, \ln \beta - 3\,\ln\Big(\frac{1 - {{\beta }^2}}{4\,\beta }\Big) 
      \,\bigg]
       \,\nonumber\,\\ 
 & & \mbox{}\,\qquad + 
     \frac{237 - 96\,\beta  + 62\,{{\beta }^2} + 32\,{{\beta }^3} - 
        59\,{{\beta }^4}}{4}\,\ln p - 
     16\,\beta \,( 3 - {{\beta }^2} ) \,\ln\Big(\frac{1 + \beta }{4}\Big)\,
      \nonumber\,\\ 
 & & \mbox{}\,\qquad - 
     2\,\beta \,( 39 - 17\,{{\beta }^2} ) \,
      \ln\Big(\frac{1 - {{\beta }^2}}{2\,{{\beta }^2}}\Big) - 
     \frac{\beta \,\left( 75 - 29\,{{\beta }^2} \right) }{2}\,\bigg\} 
\,.
\label{deltatwoloop}
\end{eqnarray}
The threshold and high energy behaviour is given by
\begin{eqnarray}
\delta^{(2)} & \stackrel{\beta\to 0}{\longrightarrow} &
  3\,\zeta(2)\,\ln \frac{\beta}{2}  + 
  \left( -\frac{3}{2} + 8\,\ln 2 \right) \,\beta + 
  2\,\zeta(2)\,\Big( \ln \frac{\beta}{2} - 2 \Big) \,{\beta^2}
 \,+\,{\cal{O}}(\beta^3)
\,,\\[2mm]
\delta^{(2)} & \stackrel{x\to 0}{\longrightarrow} &
\zeta(3) - \frac{1}{2}\,\ln 2 - \frac{23}{24} - 
  3\,\left( 2\,\ln 2 + \frac{1}{2} \right) \,x 
\nonumber\\ & & + 
  \left( -3\,\ln^2 x + \ln x\,\left( 12\,\ln 2 + \frac{7}{2} \right)  - 
     4\,\zeta(3) - 18\,\zeta(2) - 5\,\ln 2 - \frac{5}{3} \right) \,{x^2} 
\\ & & + 
  \frac{2}{27}\,\left( -108\,\ln^2 x + 
     2\,\ln x\,\left( 174\,\ln 2 + 43 \right)  - 456\,\zeta(2) + 
     188\,\ln 2 + 57 \right) \,{x^3}
 \,+\,{\cal{O}}(x^4)
\,.
\nonumber
\label{deltaspecial}
\end{eqnarray}
An explicit derivation
of eq.~(\ref{mainformula}) can be found in~\cite{HTfuture}, 
where the two- and
four-body cuts are calculated separately.
\par
The complete
information on the vacuum polarization 
relevant for the internal photon
line is encoded in the moments $R_\infty^x$ and $R_0^x$. 
The crucial ingredient which allows to arrive at this simple and
compact formulation is the following condition on the high energy
behaviour of $R_{f\bar{f}}(s)$: it has to approach a constant value
$R_\infty^x$ in the limit of large $\sqrt{s}$ fast enough, which is
equivalent to the
occurrence of at most one single logarithm $\ln(-q^2/4m^2)$ in the
vacuum polarization function.
Thus we
can determine without any effort the second order correction due to a
light pair of unit-charged scalar particles with mass $m$. 
The corresponding moments for the vacuum polarization,
$\Pi_{light}^s$, as defined in (\ref{vacuumpol}) read
\begin{eqnarray}
R_\infty^{s} & = & R_{s s^*}(\infty) \, = \, \frac{1}{4},
\nonumber \\
R_0^{s} & = & 
  \int\limits_{4\,m^2}^{\infty}\,\frac{d s}{s}\,
         \left[R_{s s^*}(s)- R_\infty^s \right]
\, = \,
-\frac{2}{3}+\frac{1}{4} \ln 4
\,.
\label{momentascalar}
\end{eqnarray}

\begin{figure}[ht]
 \begin{center}
 \begin{tabular}{ccccc}
   \epsfxsize=2.8cm
   \leavevmode
   \epsffile[150 270 470 520]{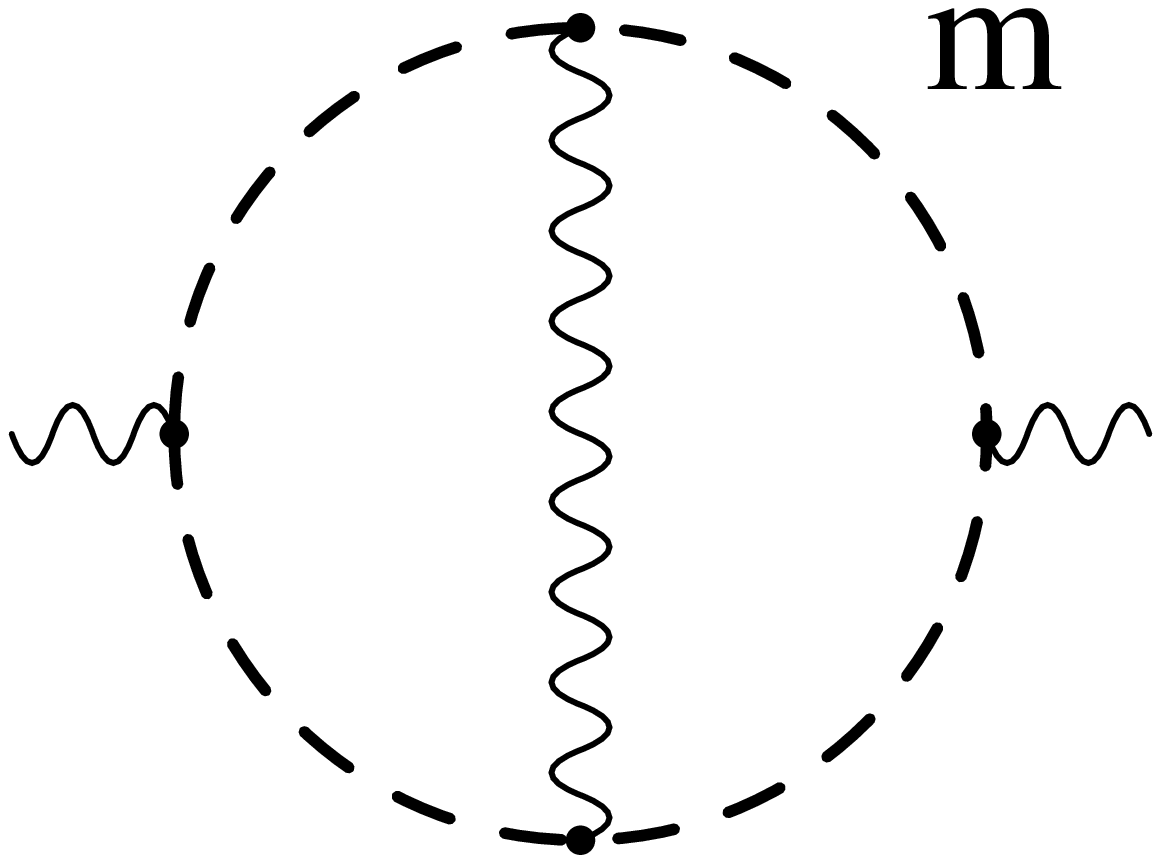}
   &
   \epsfxsize=2.8cm
   \leavevmode
   \epsffile[150 270 470 520]{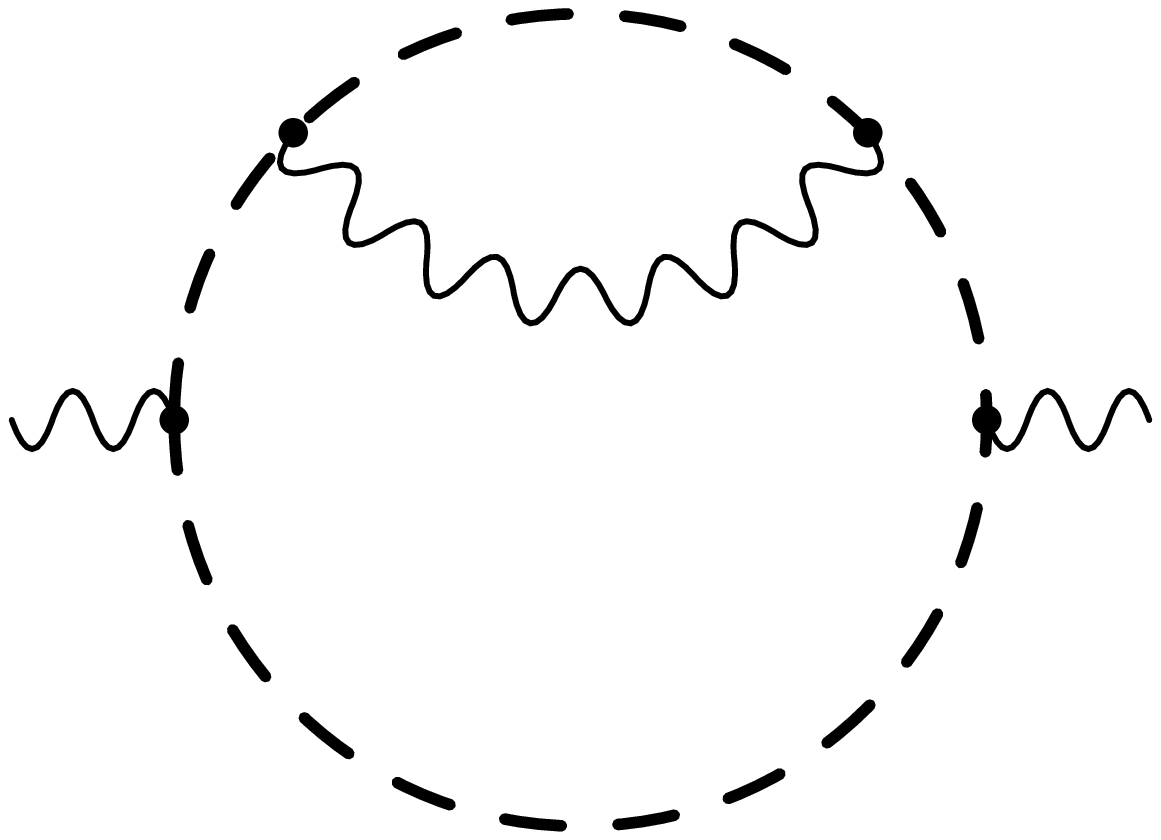}
   &
   \epsfxsize=2.8cm
   \leavevmode
   \epsffile[150 270 470 520]{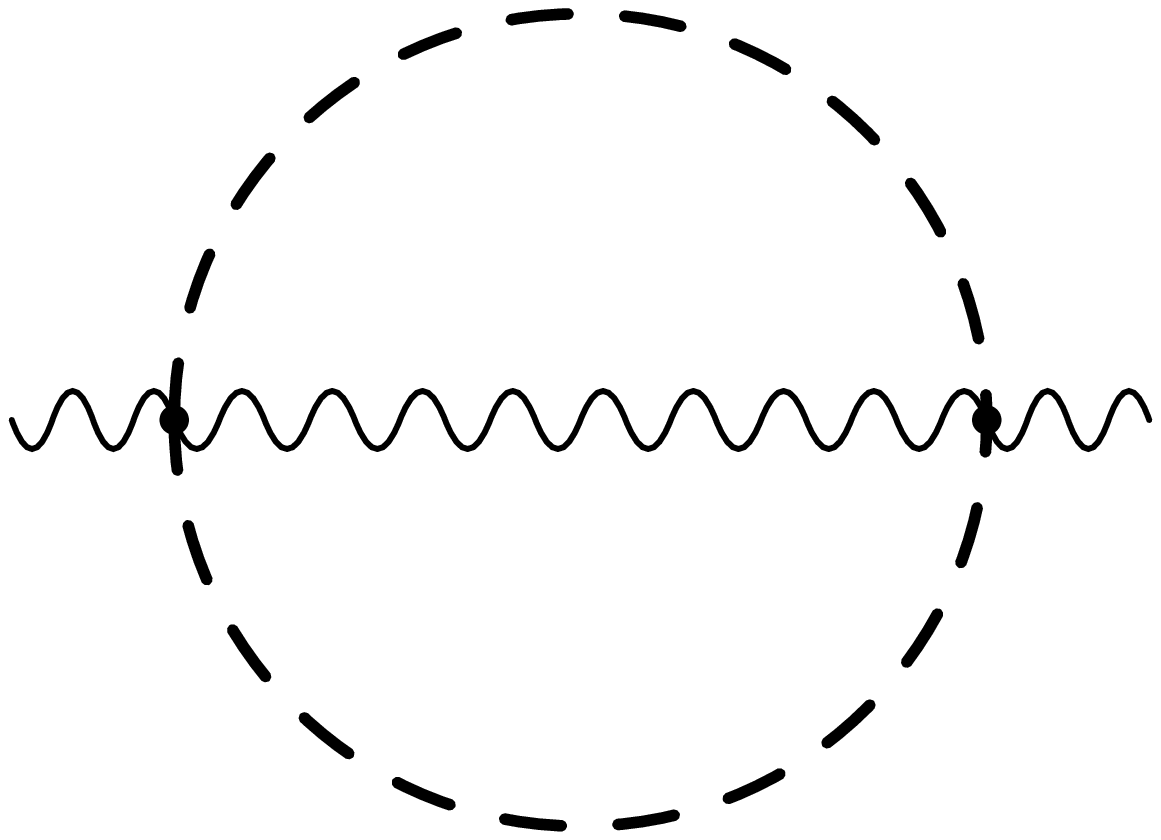}
   &
   \epsfxsize=2.8cm
   \leavevmode
   \epsffile[150 270 470 520]{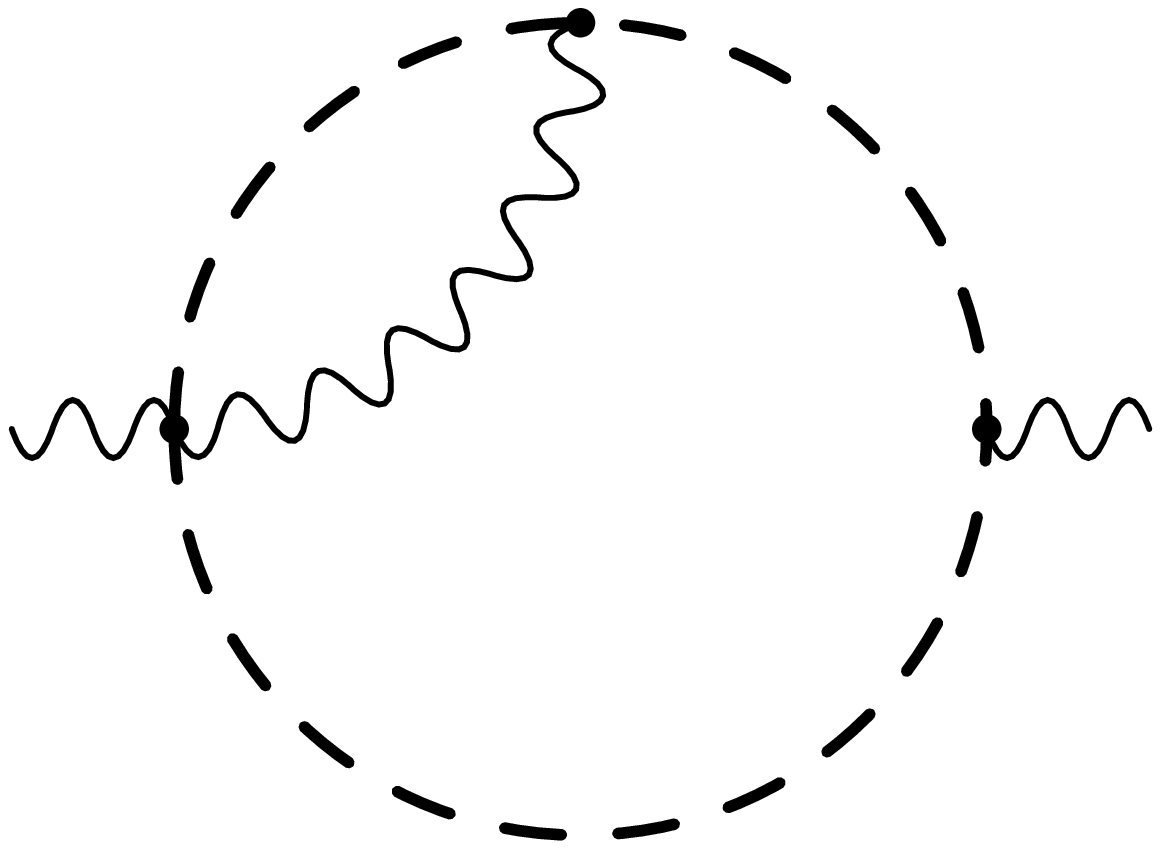}
   &
   \epsfxsize=2.8cm
   \leavevmode
   \epsffile[150 270 470 520]{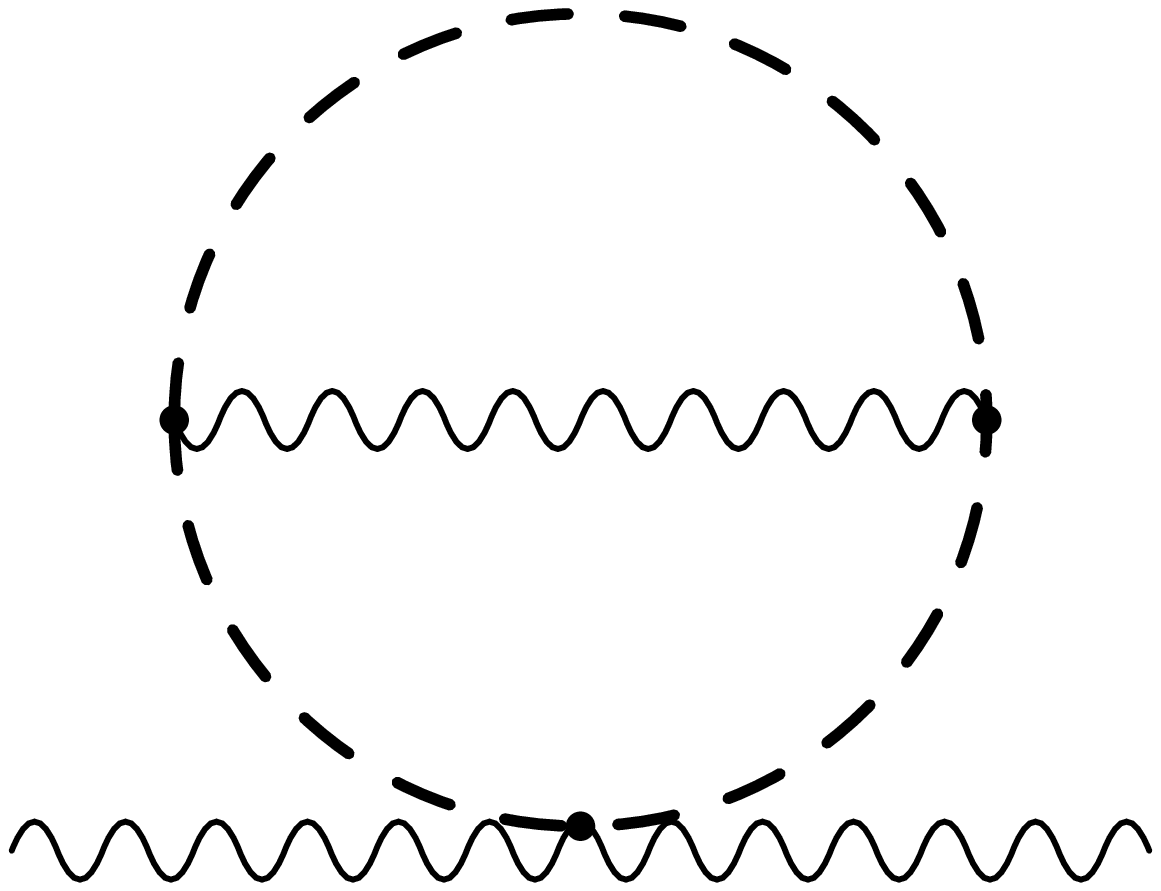}
 \end{tabular}
 \caption{\label{fig2lscalar} Scalar two-loop diagrams to the  
      vacuum polarization. The dashed line represents a scalar particle 
      with mass $m$.}
 \end{center}
\end{figure}

Because the ${\cal{O}}(\alpha)$ corrections to the cross-section
$R_{f\bar f}$ also approach a constant value for high energies even
${\cal{O}}(\alpha^3)$ corrections to $R_{F\bar F}$ due to a light
fermion pair with additional real and virtual radiation of a photon off
the light fermions can be calculated. 
These ${\cal O}(\alpha^3)$ contributions can be cast into the form
of eq.~(\ref{mainformula}) with the following two moments from the 
vacuum polarization function $\Pi_{light}^{f\gamma}$, 
\begin{eqnarray}
R_\infty^{f\,\gamma} & = & \frac{\alpha}{\pi} \frac{3}{4}
\,,\nonumber \\
R_0^{f\,\gamma} & = & \frac{\alpha}{\pi}\left(
            -\frac{5}{8}+3\zeta(3)
            +\frac{3}{4}\ln4
                                        \right) 
\,.
\label{momentafermionphoton}
\end{eqnarray}
Similar arguments hold for the ${\cal{O}}(\alpha)$
corrections to $R_{s s^*}$. Here the moments read
\begin{eqnarray}
R_\infty^{s\,\gamma} & = & \frac{\alpha}{\pi} \frac{3}{4}
\,,\nonumber \\
R_0^{s\,\gamma} & = & \frac{\alpha}{\pi}\left(
            -\frac{49}{16}+\frac{3}{4}\zeta(3)
            +\frac{3}{4}\ln4
                                        \right) 
\,,
\label{momentasclarphoton}
\end{eqnarray}
where the vacuum polarization diagrams depicted in 
Fig.~\ref{fig2lscalar} have to be taken into account.

\vspace{.3cm}

\vspace{.3cm}
\par
The mass singularity in the limit $m\to0$, evident from 
eqs.~(\ref{mainformula}) and (\ref{vacuumpol}) is
a consequence of the renormalization scheme 
with the fine structure constant $\alpha$ 
defined at momentum transfer zero.
For the vacuum polarization due to massless
particles a mass
independent renormalization scheme like
$\overline{\mbox{MS}}$ for the coupling is more
appropriate. 
The ${\cal{O}}(\alpha^2)$ relation between the fine structure constant and
the running ${\overline{\mbox{MS}}}$ coupling at the scale $\mu$ for 
the case of either a light fermion or a light scalar reads
\begin{equation}
\alpha \, = \, \alpha_{\overline{\mbox{\tiny MS}}}(\mu^2) \,
         \left(\,1+\frac{\alpha_{\overline{\mbox{\tiny MS}}}(\mu^2)}{\pi}\,
                   \frac{1}{3}\,R_\infty^x\,\ln\frac{m^2}{\mu^2}\,\right)
          + {\cal O}(\alpha_{\overline{\mbox{\tiny MS}}}^3)
\,.
\label{alphaqedtorun}
\end{equation}
Replacing the fine structure constant in eq.~(\ref{rexpanded}) by
$\alpha_{\overline{\mbox{\tiny MS}}}$ results in
\begin{equation}
R_{F\bar F}\, = \, r^{(0)} + 
\,\Big(\frac{\alpha_{\overline{\mbox{\tiny MS}}}(\mu^2)}
           {\pi}\Big)\,r^{(1)} +
\,\Big(\frac{\alpha_{\overline{\mbox{\tiny MS}}}(\mu^2)}
           {\pi}\Big)^2\,r^{(2)}_{\overline{\mbox{\tiny MS}}} +\, ...\,,
\end{equation}
where
\begin{equation}
r^{(2)}_{x,\overline{\mbox{\tiny MS}}} \, = \,
\left\{\,
-\frac{1}{3}\Big[\,R_\infty^{x,{\overline{\mbox{\tiny MS}}}}\,
 \ln\frac{\mu^2}{s} - R_0^{x,{\overline{\mbox{\tiny MS}}}}\,\Big]\,
 r^{(1)} +
R_\infty^{x,{\overline{\mbox{\tiny MS}}}}\,\delta^{(2)}  
  \,\right\}\,,\qquad x\,=\,f,s.
\label{r2msbar}
\end{equation}
It is evident that eq.~(\ref{r2msbar}) closely resembles 
eq.~(\ref{mainformula}).
Also in the $\overline{\mbox{MS}}$ scheme the moments are uniquely
determined via the vacuum polarization function
\begin{eqnarray}
\Pi_{massless}^{x,\overline{\mbox{\tiny MS}}}(q^2)  
         &=&
 -\frac{\alpha_{\overline{\mbox{\tiny MS}}}(\mu^2)}{3\,\pi}\,\Big[\,
   R_\infty^{x,\overline{\mbox{\tiny MS}}}\,
   \ln\frac{-q^2}{4\,\mu^2} + R_0^{x,\overline{\mbox{\tiny MS}}}
   \,\Big]\,,
\label{vacuumpolmsbar}
\end{eqnarray}
where
\begin{equation}
R_\infty^{x,\overline{\mbox{\tiny MS}}}=R_\infty^x\,,\,\,
R_0^{x,\overline{\mbox{\tiny MS}}}=R_0^x\,\,\,\,
\mbox{for}\,\,\,\,\, x=f, s.
\end{equation}
It should be noted that the equality of the zero-moments, $R_0^x$,
in the $\overline{\mbox{MS}}$ and on-shell scheme holds because there
are no non-logarithmic terms present in eq.~(\ref{alphaqedtorun}).
For the three-loop corrections which were mentioned above again
the r.h.s. of eq.~(\ref{r2msbar}) can be employed in the 
$\overline{\mbox{MS}}$ scheme. The corresponding $\overline{\mbox{MS}}$
moments read
\begin{eqnarray}
R_\infty^{f\,\gamma,\overline{\mbox{\tiny MS}}} 
& = & \frac{\alpha_{\overline{\mbox{\tiny MS}}}}{\pi} \frac{3}{4}
\,,\nonumber \\
R_0^{f\,\gamma,\overline{\mbox{\tiny MS}}} & = & 
    \frac{\alpha_{\overline{\mbox{\tiny MS}}}}{\pi} \left(-\frac{55}{16}
                  +3\zeta(3)
                  +\frac{3}{4}\ln 4
                         \right)
\,,\nonumber \\
R_\infty^{s\,\gamma,\overline{\mbox{\tiny MS}}} 
& = & \frac{\alpha_{\overline{\mbox{\tiny MS}}}}{\pi} \frac{3}{4}
\,,\nonumber \\
R_0^{s\,\gamma,\overline{\mbox{\tiny MS}}} & = & 
    \frac{\alpha_{\overline{\mbox{\tiny MS}}}}{\pi} \left(-\frac{43}{16}
                  +\frac{3}{4}\zeta(3)
                  +\frac{3}{4}\ln 4
                         \right)
\,.
\label{momentazeromsbar}
\end{eqnarray}
Here, the zero-moments in the 
$\overline{\mbox{MS}}$ scheme are different from to the corresponding
ones in the on-shell scheme, 
eqs.~(\ref{momentafermionphoton},\ref{momentasclarphoton}), 
due to a non-logarithmic contribution in the ${\cal O}(\alpha^3)$
relation between the fine structure constant and 
$\alpha_{\overline{\mbox{\tiny MS}}}$ ($x=f\gamma, s\gamma$):
\begin{eqnarray}
\alpha \, = \, \alpha_{\overline{\mbox{\tiny MS}}}(m^2) \,
           \left[\,1 + \frac{1}{3}
  \left(\frac{\alpha_{\overline{\mbox{\tiny MS}}}(m^2)}{\pi}\right)\,
  \left( R_0^{x,\overline{\mbox{\tiny MS}}} - R_0^x \right)
    \,\right]
          + {\cal O}(\alpha_{\overline{\mbox{\tiny MS}}}^4)
\,.
\end{eqnarray}
Eqs.~(\ref{r2msbar}) and (\ref{vacuumpolmsbar}) provide a simple and 
unambiguous method to determine second order corrections to $R_{F\bar
F}$ due to the vacuum polarization from arbitrary massless
particles. For that one has to compute the 
$\overline{\mbox{MS}}$-renormalized one-loop vacuum polarization
function, $\Pi_{massless}^{x,\overline{\mbox{\tiny MS}}}$, where the
index $x$ denotes the types of massless particles considered, and to
identify the moments 
$R_\infty^{x,\overline{\mbox{\tiny MS}}}$ and 
$R_0^{x,\overline{\mbox{\tiny MS}}}$, as defined in
eq.~(\ref{vacuumpolmsbar}). Inserting the moments into
eq.~(\ref{r2msbar}) gives the desired result.
\par
\vspace{1cm}\noindent
\section{Transition to QCD and Gluon Bubble Contribution}
\label{sectiontransition}
To obtain the ${\cal{O}}(\alpha_s^2)$ corrections to massive quark 
production due to
massless quarks (or squarks) and the ${\cal{O}}(\alpha_s^3)$ contributions
corresponding to the ${\cal{O}}(\alpha^3)$ corrections presented in the 
previous section 
we have to multiply the QED results by the
corresponding SU(3) group theoretical factors
$T=1/2$ (the moments 
$R_\infty^{x,\overline{\mbox{\tiny MS}}}$, 
$R_0^{x,\overline{\mbox{\tiny MS}}}$, 
$R_\infty^{x\,\gamma,\overline{\mbox{\tiny MS}}}$,  and 
$R_{0}^{x\,\gamma,\overline{\mbox{\tiny MS}}}$ $(x=f,s)$) and 
$C_F=4/3$ (the $r^{(1)}$, 
$\delta^{(2)}$,
$R_\infty^{x\,\gamma,\overline{\mbox{\tiny MS}}}$,  and 
$R_{0}^{x\,\gamma,\overline{\mbox{\tiny MS}}}$ $(x=f,s)$).
For the ${\cal{O}}(\alpha_s^3)$ contributions 
this leads to the colour factor $T\,C_F^2$.
Furthermore an additional global colour factor $N_c=3$ has to be 
taken into account.
$\alpha_{\overline{\mbox{\tiny MS}}}$ now represents 
the $\overline{\mbox{MS}}$ renormalized QCD coupling constant.
\par
\begin{figure}[ht]
 \begin{center}
 \begin{tabular}{ccc}
   \epsfxsize=3.5cm
   \leavevmode
   \epsffile[170 270 420 520]{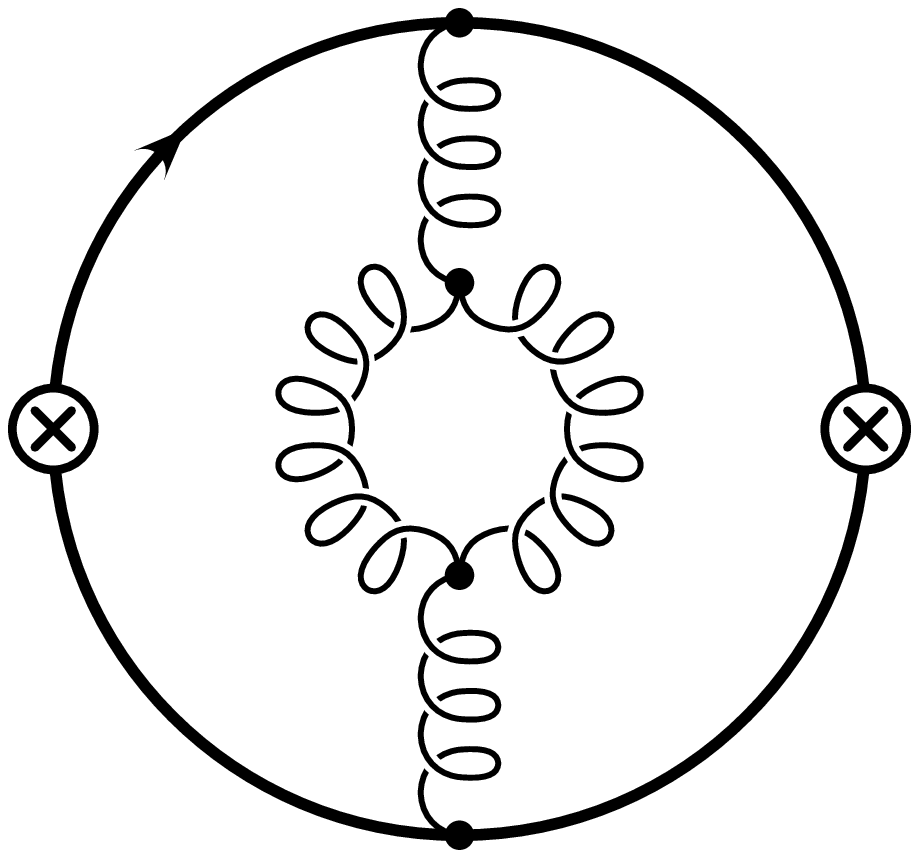}
   & \hspace{10ex} &
   \epsfxsize=3.5cm
   \leavevmode
   \epsffile[170 270 420 520]{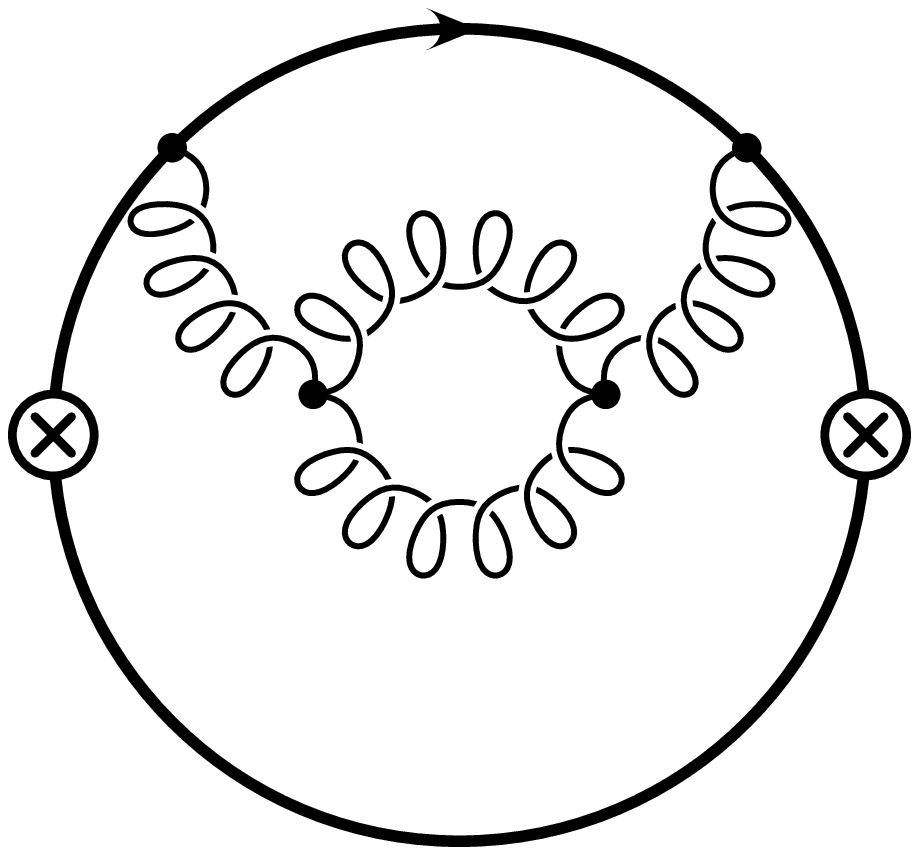}
 \end{tabular}
 \caption{\label{figgluons} Gluonic double bubble diagrams. The
    ghost diagrams are not depicted explicitely.}
 \end{center}
\end{figure}

Using the method described in section~2 we are now in a 
position to determine the gluonic self energy
contribution to $r^{(2)}$, as illustrated 
in Fig.~\ref{figgluons}, by determining the
corresponding moments for the gluonic contributions to the 
${\cal{O}}(\alpha_s)$ gluon propagator. The moments
which correspond to eq.~(\ref{vacuumpolmsbar}) read
\begin{eqnarray}
R_\infty^{g,\overline{\mbox{\tiny MS}}} 
& = & C_A\left(-\frac{5}{4} - \frac{3}{8}\xi\right),
\nonumber \\
R_0^{g,\overline{\mbox{\tiny MS}}} 
& = & C_A\left(\frac{31}{12}-\frac{3}{4}\xi+\frac{3}{16}\xi^2
                    +\left(-\frac{5}{4}-\frac{3}{8}\xi\right)\ln 4\right)
\,,
\end{eqnarray}
where the gauge parameter $\xi$ is defined via the gluon propagator
in lowest order
\begin{equation}
\frac{i}{q^2+i\,\epsilon}\,\left(
-\,g^{\mu\nu} + \xi\,\frac{q^\mu\,q^\nu}{q^2}
\right)
\,.
\end{equation}
Of course, the moments and $r_{g,\overline{\mbox{\tiny MS}}}^{(2)}$ 
are not gauge invariant. 
However, it is an
interesting fact that for the special choice $\xi=4$ the combination
$\alpha_s^2\,R_\infty^{g,\overline{\mbox{\tiny MS}}}/(3\,\pi)$
coincides with the gluonic contribution to the 
QCD $\beta$-function of ${\cal{O}}(\alpha_s^2)$. 
Thus for $\xi=4$ the term $r_{g,\overline{\mbox{\tiny MS}}}^{(2)}$ 
accounts for the leading 
logarithmic behaviour of the sum of all gluonic ${\cal{O}}(\alpha_s^2)$ 
diagrams in the high energy limit.
It is quite obvious that such a $\xi$ can be found.
Remarkably enough, the complete gluonic contributions of 
${\cal O}(\alpha_s^2)$ to the QCD potential are described by
$r_{g,\overline{\mbox{\tiny MS}}}^{(2)}$ 
for the choice $\xi=4$. To be specific,
\begin{eqnarray}
 V_{\mbox{\scriptsize QCD}}(Q^2) &=& 
         -4\pi C_F\frac{\alpha_V(Q^2)}{Q^2}\
\end{eqnarray}
with \cite{Fis77} 
\begin{eqnarray}
 \alpha_V(Q^2) &=& \alpha_s(\mu^2)
     \left(1-\Pi^{g,\overline{\mbox{\tiny MS}}}(-Q^2)|_{\xi=4}\right)
\nonumber\\
&=&  
  \alpha_s(\mu^2)\Bigg[
      1 + \frac{\alpha_s(\mu^2)}{3\pi} C_A \left( 
          -\frac{11}{4}\ln\frac{Q^2}{\mu^2}+\frac{31}{12}
                                           \right)
\Bigg]
\,.
\label{alphav}
\end{eqnarray}
For this choice of gauge the leading threshold behaviour of 
order $\alpha_s^2$ with the colour structure proportional
to $C_AC_F$ is incorporated in the gluonic double bubble
diagrams. This can be seen as follows:
The leading term to $R_{F\bar F}$ of ${\cal O}(\alpha_s)$ is given by
\begin{eqnarray}
R_{F\bar{F}} & \stackrel{\beta\to 0}{\longrightarrow} &
            N_c\,C_F\frac{3\,\pi}{4}\,\alpha_s
\,.
\label{rffthreshold}
\end{eqnarray}
(The leading term of order $\alpha_s^2$ which is proportional to
$\pi^2/\beta$ can be derived from Sommerfeld's rescattering
formula for the Coulomb problem. It is, however, proportional
to $C_F^2$.)
We are interested in terms of order $\alpha_s^2$ proportional 
to $C_AC_F$ which are
characteristic for the non-abelian nature of the 
interaction. On the basis of general considerations this term
is obtained from eq.~(\ref{rffthreshold}) through the
replacement of $\alpha_s$ by $\alpha_V(\beta^2 s)$ as given
in eq.~(\ref{alphav}). This prediction of the leading
logarithmic and constant $C_AC_F$ term coincides with the
result of the analytic calculation of the diagrams 
(Fig. \ref{figgluons}) in the $\xi=4$ gauge. This aspect
strongly resembles the behaviour of the leading
threshold contribution from massless quark loops as discussed
in 
\cite{HoaKueTeu95_2BroHoaKueTeu95}
where a similar relation between the leading 
threshold terms  and eq.~(\ref{rffthreshold}) expressed through
an effective coupling constant has been observed.
\section{Application at ${\cal{O}}(\alpha_s^4)$} 
\label{sectionapplication}
\begin{figure}[t]
 \begin{center}
 \begin{tabular}{cccc}
   \epsfxsize=2.8cm
   \leavevmode
   \epsffile[150 270 470 520]{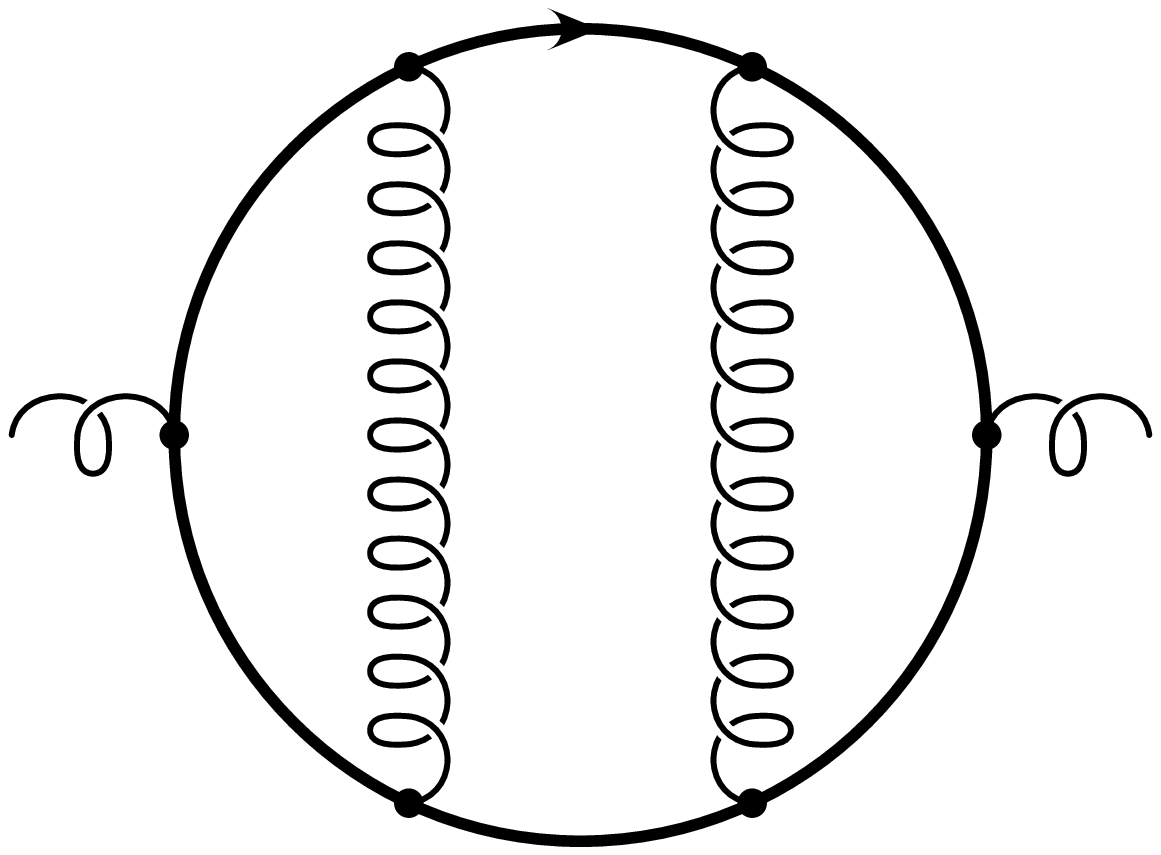}
   &
   \epsfxsize=2.8cm
   \leavevmode
   \epsffile[150 270 470 520]{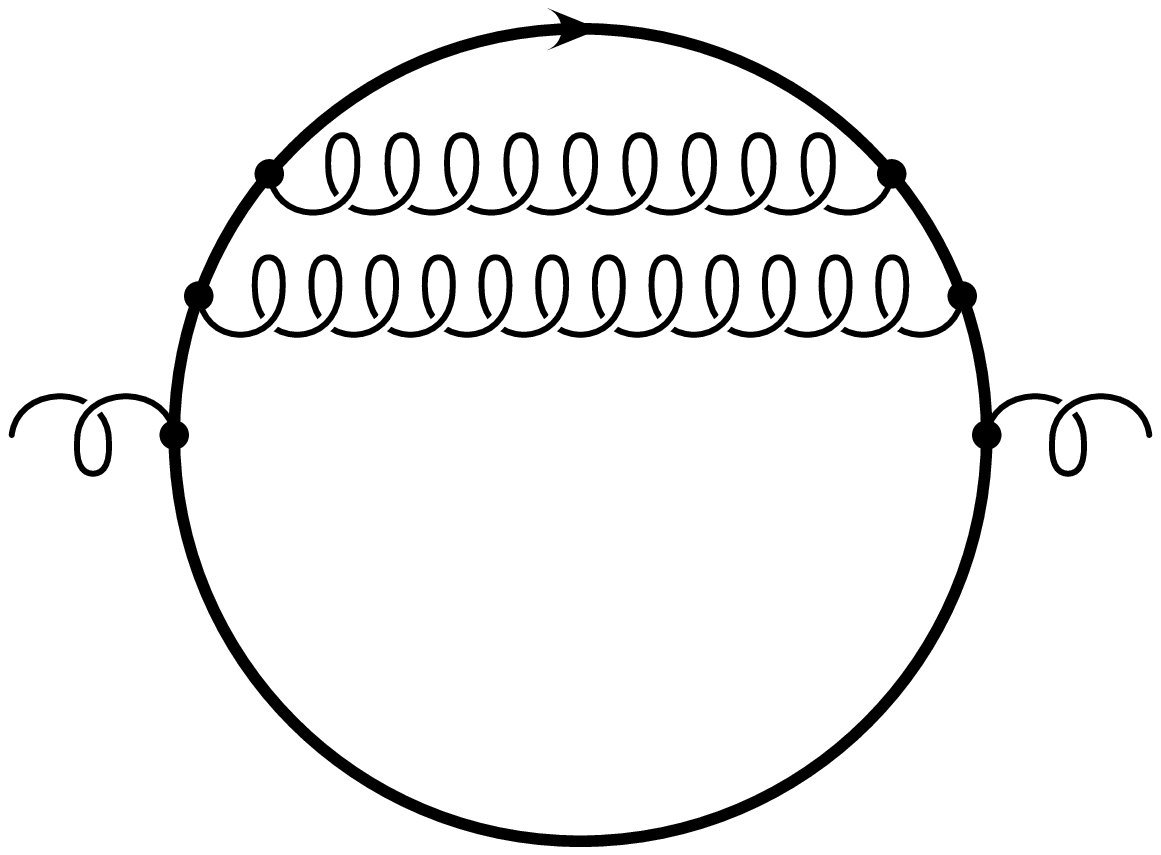}
   &
   \epsfxsize=2.8cm
   \leavevmode
   \epsffile[150 270 470 520]{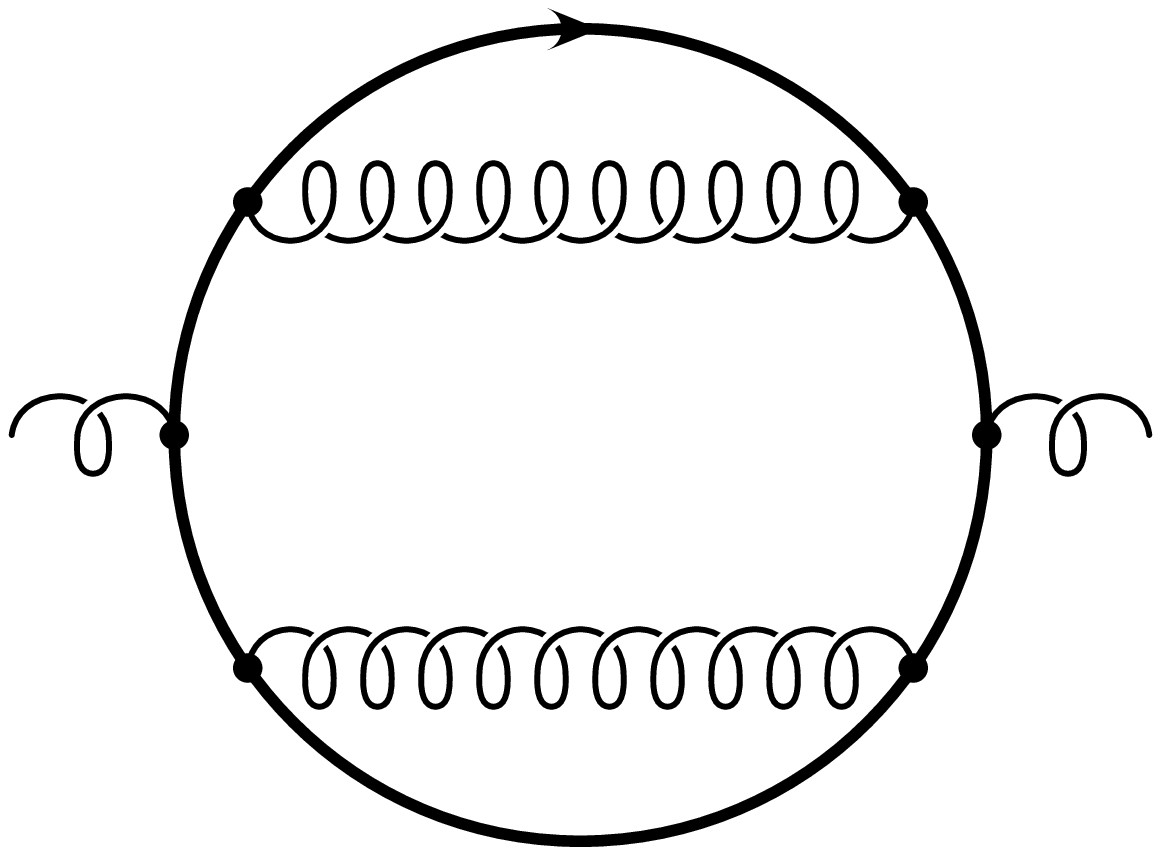}
   &
   \epsfxsize=2.8cm
   \leavevmode
   \epsffile[150 270 470 520]{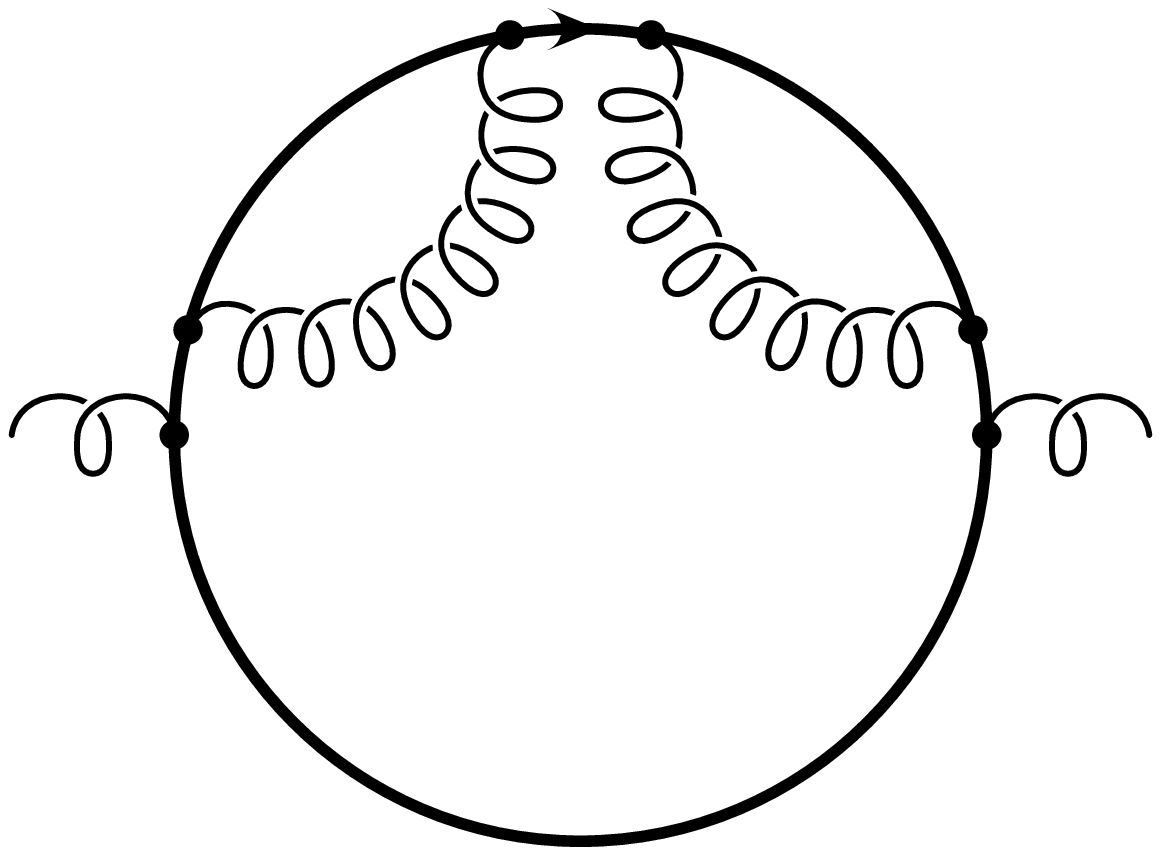}
\\
   \epsfxsize=2.8cm
   \leavevmode
   \epsffile[150 270 470 520]{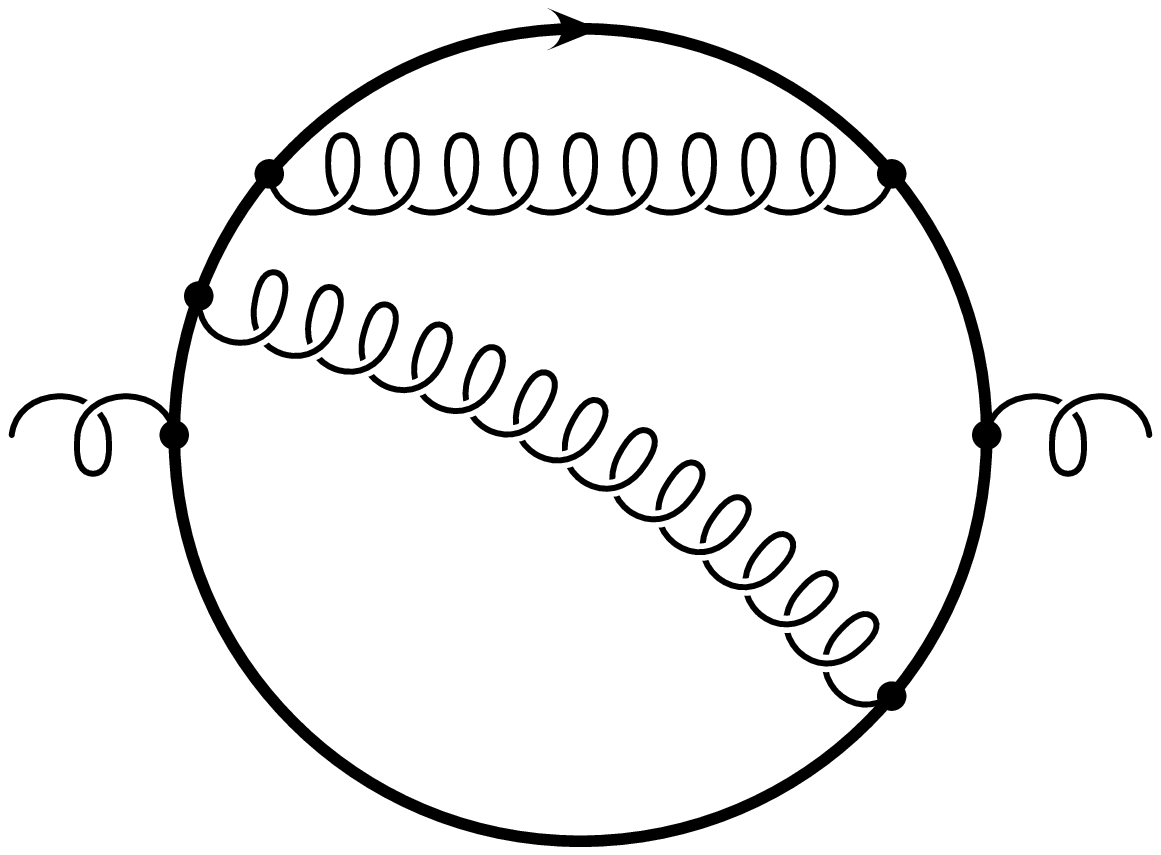}
   &
   \epsfxsize=2.8cm
   \leavevmode
   \epsffile[150 270 470 520]{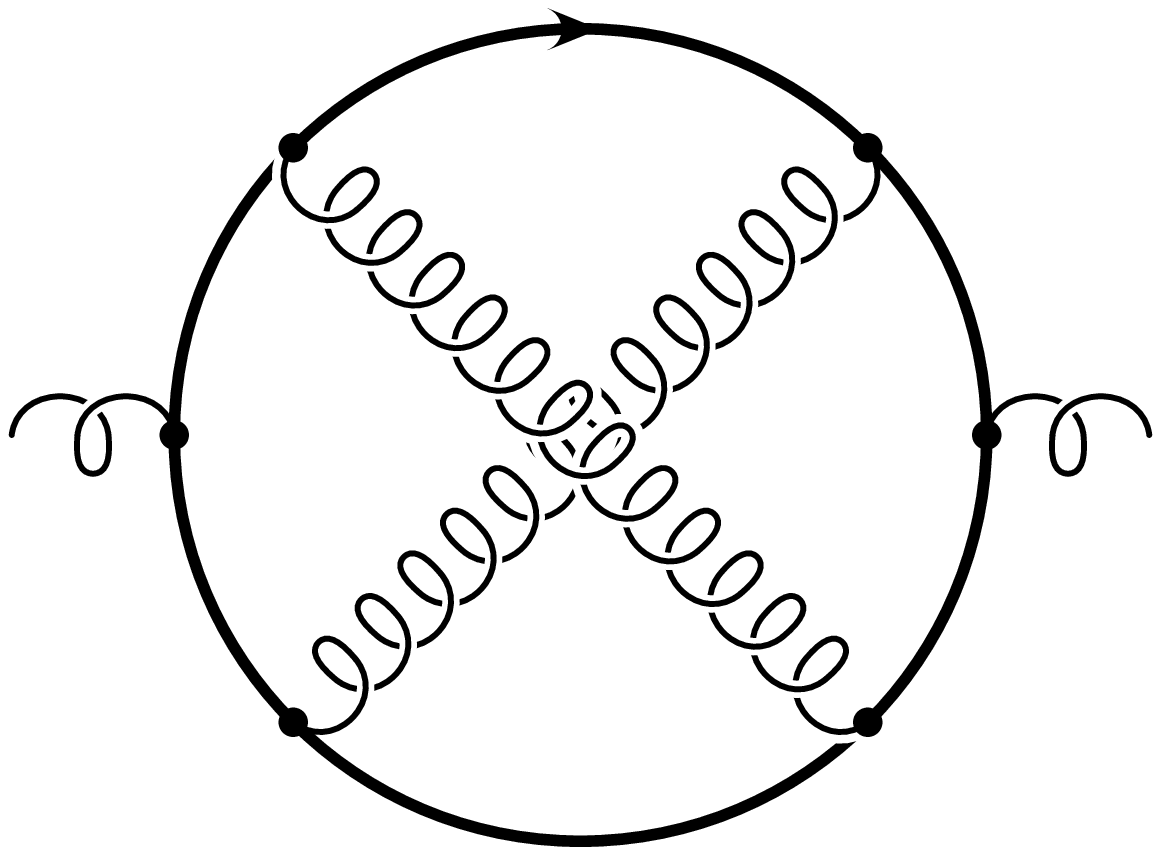}
   &
   \epsfxsize=2.8cm
   \leavevmode
   \epsffile[150 270 470 520]{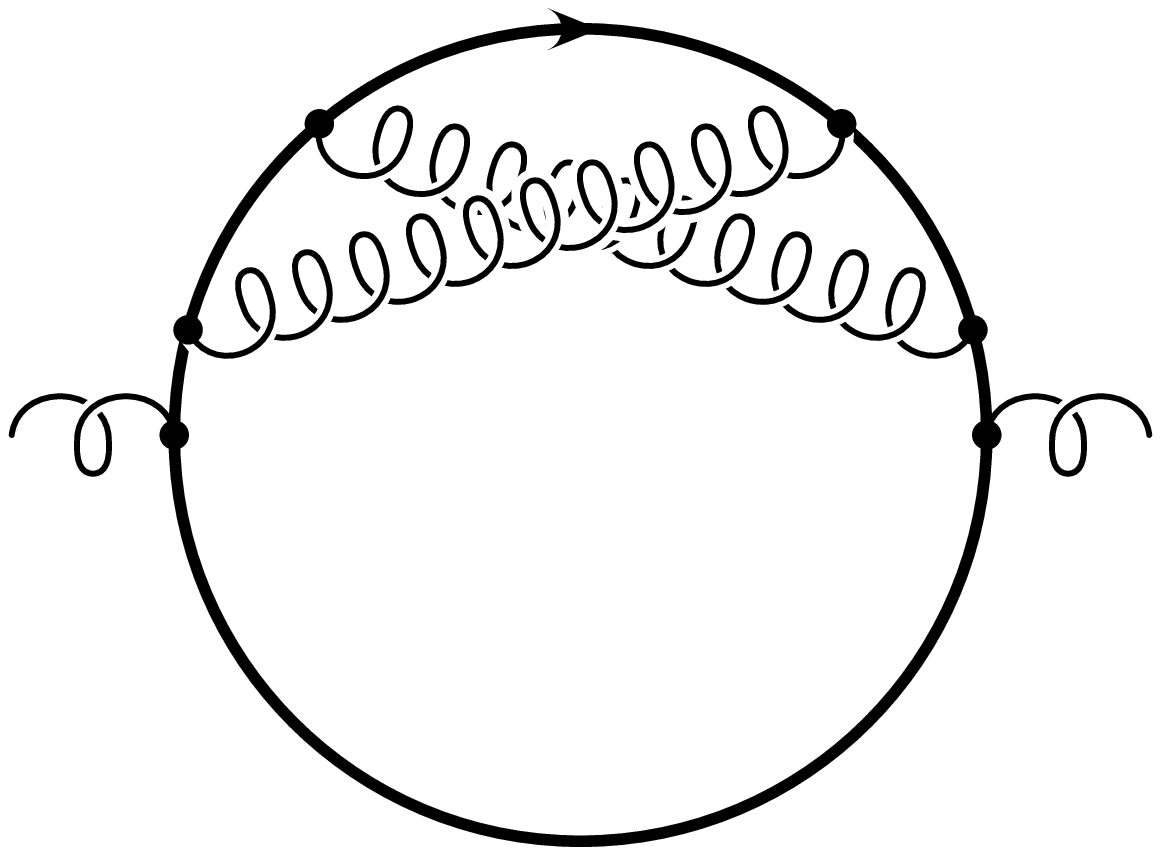}
   &
   \epsfxsize=2.8cm
   \leavevmode
   \epsffile[150 270 470 520]{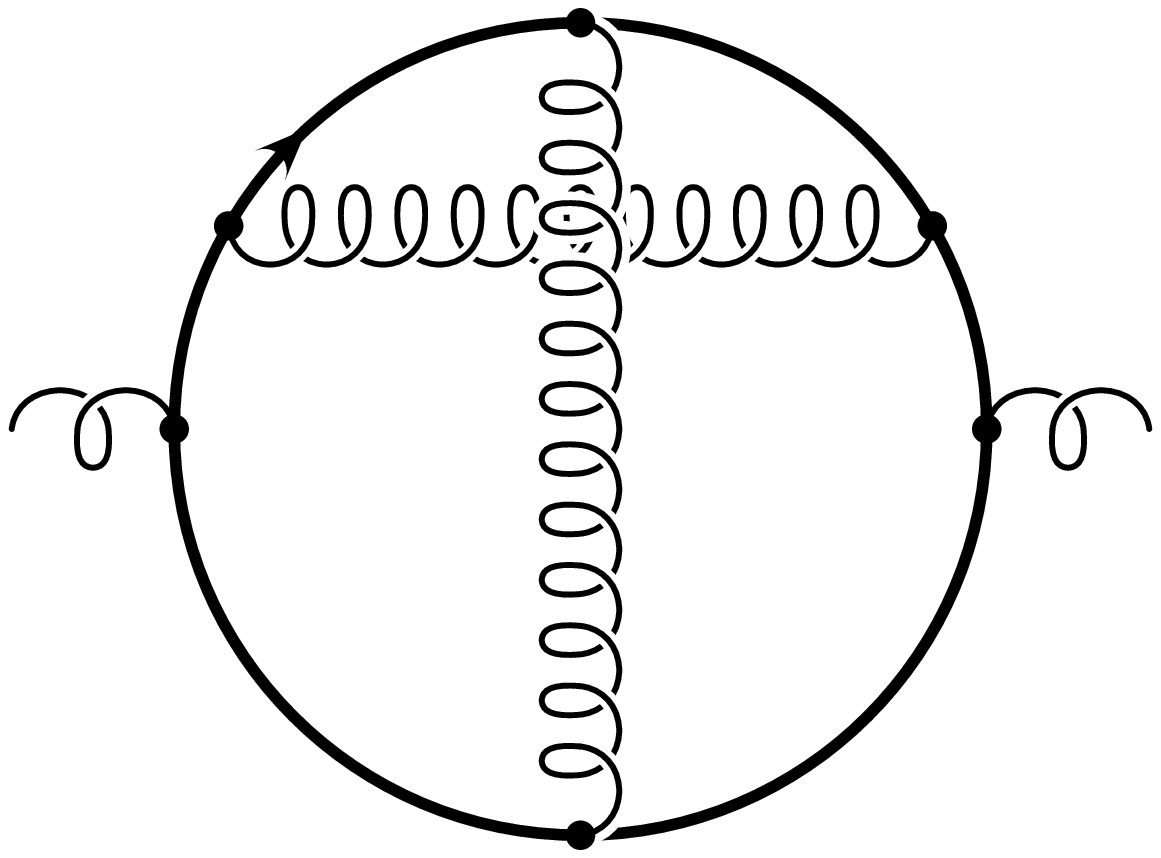}
 \end{tabular}
 \caption{\label{fig3lgl} Three-loop diagrams contributing to
                          $R_\infty^{fgg,\overline{\mbox{\tiny MS}}}$
                          and $R_0^{fgg,\overline{\mbox{\tiny MS}}}$.}
 \end{center}
\end{figure}

The concept of the moments allows for the determination of higher order
QCD corrections from one gluon exchange diagrams with the insertion of
massless vacuum polarization functions into the gluon line
which contain at most one single
logarithm $\ln(-q^2/4\mu^2)$. This condition is fulfilled by the 
sum of those terms in the
third order vacuum polarization due to a massless fermion 
anti-fermion pair (depicted in Fig.~\ref{fig3lgl}) which are
proportional to $T\,C_F^2$. It is the same class of contributions
which would also be present in an abelian theory.
The result for the corresponding vacuum polarization function in the
$\overline{\mbox{MS}}$ scheme reads 
\cite{GorKatLar91}
\begin{eqnarray}
\Pi_{massless}^{fgg,\overline{\mbox{\tiny MS}}}(q^2)  
         & = &
 -\frac{\alpha_s(\mu^2)}{3\,\pi}\,\Big[\,
   R_\infty^{fgg,\overline{\mbox{\tiny MS}}}\,
   \ln\frac{-q^2}{4\,\mu^2} + R_0^{fgg,\overline{\mbox{\tiny MS}}}
   \,\Big]\,,
\label{vacuumpolfgg}
\end{eqnarray}
with
\begin{eqnarray}
R_\infty^{fgg,\overline{\mbox{\tiny MS}}} & = &
\left(\frac{\alpha_s(\mu^2)}{\pi}\right)^2 \,C_F^2\, T \,
   \left(-\frac{3}{32}\right)\,,
\nonumber\\[2mm]
R_0^{fgg,\overline{\mbox{\tiny MS}}} & = &
\left(\frac{\alpha_s(\mu^2)}{\pi}\right)^2 \,C_F^2\, T \,\left(
    \frac{143}{96} + \frac{37}{8}\zeta(3) - \frac{15}{2}\zeta(5)   
   -\frac{3}{32} \ln4 \right)
\,.
\end{eqnarray}
The same is also true for the corresponding
third order contributions to the vacuum polarization 
due to a massless squark anti-squark pair, 
$\Pi_{massless}^{sgg,\overline{\mbox{\tiny MS}}}$
with the moments
\cite{BroDelKre96}
\begin{eqnarray}
R_\infty^{sgg,\overline{\mbox{\tiny MS}}} & = &
\left(\frac{\alpha_s(\mu^2)}{\pi}\right)^2 \,C_F^2\, T \,
   \frac{87}{128}\,,
\nonumber\\[2mm]
R_0^{sgg,\overline{\mbox{\tiny MS}}} & = &
\left(\frac{\alpha_s(\mu^2)}{\pi}\right)^2 \,C_F^2\, T \,\left(
   -\frac{251}{96} + \frac{5}{2}\zeta(3) - \frac{15}{8}\zeta(5)   
   +\frac{87}{128} \ln4 \right)
\,.
\end{eqnarray}
Thus the corrections to the total heavy quark production cross-section
from the insertion of $\Pi_{massless}^{xgg,\overline{\mbox{\tiny MS}}}$
into the one gluon exchange diagrams read ($x=f$ for massless fermions 
or $x=s$ for massless squarks)
\begin{equation}
r^{(4)}_{xgg,\overline{\mbox{\tiny MS}}} \, = \,
N_c\,\left\{\,
-\frac{1}{3}\Big[\,R_\infty^{xgg,{\overline{\mbox{\tiny MS}}}}\,
 \ln\frac{\mu^2}{s} - R_0^{xgg,{\overline{\mbox{\tiny MS}}}}\,\Big]\,
 C_F\,r^{(1)} +
R_\infty^{xgg,{\overline{\mbox{\tiny MS}}}}\,C_F\,\delta^{(2)}  
  \,\right\}\,.
\label{r4fgg}
\end{equation}
This four-loop result exemplifies the power of the concept of
moments used in this paper to evaluate massive higher order contributions
to the production of heavy quarks.
\newpage
\sloppy
\raggedright
\def\app#1#2#3{{\it Act. Phys. Pol. }{\bf B #1} (#2) #3}
\def\apa#1#2#3{{\it Act. Phys. Austr.}{\bf #1} (#2) #3}
\def\lhc{Proc. LHC Workshop, CERN 90-10}
\def\npb#1#2#3{{\it Nucl. Phys. }{\bf B #1} (#2) #3}
\def\plb#1#2#3{{\it Phys. Lett. }{\bf B #1} (#2) #3}
\def\prd#1#2#3{{\it Phys. Rev. }{\bf D #1} (#2) #3}
\def\pR#1#2#3{{\it Phys. Rev. }{\bf #1} (#2) #3}
\def\prl#1#2#3{{\it Phys. Rev. Lett. }{\bf #1} (#2) #3}
\def\prc#1#2#3{{\it Phys. Reports }{\bf #1} (#2) #3}
\def\cpc#1#2#3{{\it Comp. Phys. Commun. }{\bf #1} (#2) #3}
\def\nim#1#2#3{{\it Nucl. Inst. Meth. }{\bf #1} (#2) #3}
\def\pr#1#2#3{{\it Phys. Reports }{\bf #1} (#2) #3}
\def\sovnp#1#2#3{{\it Sov. J. Nucl. Phys. }{\bf #1} (#2) #3}
\def\jl#1#2#3{{\it JETP Lett. }{\bf #1} (#2) #3}
\def\jet#1#2#3{{\it JETP Lett. }{\bf #1} (#2) #3}
\def\zpc#1#2#3{{\it Z. Phys. }{\bf C #1} (#2) #3}
\def\ptp#1#2#3{{\it Prog.~Theor.~Phys.~}{\bf #1} (#2) #3}
\def\nca#1#2#3{{\it Nouvo~Cim.~}{\bf #1A} (#2) #3}

\end{document}